\documentclass[oldversion]{aa}

\usepackage{graphicx}
\usepackage{txfonts}
\begin{document}

\title{8.4\,GHz VLBI observations of SN\,2004et in NGC\,6946.}

\author{I.~Mart\'{\i}-Vidal\inst{1} \and J.~M.~Marcaide\inst{1} \and A.~Alberdi\inst{2} \and
        J.~C.~Guirado\inst{1} \and L.~Lara\inst{2}\fnmsep\thanks{Deceased} \and
        M.~A.~P\'erez-Torres\inst{2} \and E.~Ros\inst{3} \and M.~K.~Argo\inst{4} \and
        R.~J.~Beswick\inst{4} \and T.~W.~B.~Muxlow\inst{4} \and A.~Pedlar\inst{4} \and
        I.~I.~Shapiro\inst{5} \and C.~J.~Stockdale\inst{6} \and R.~A.~Sramek\inst{7} \and
        K.~W.~Weiler\inst{8} \and J.~Vinko\inst{9}}

\offprints{I. Marti-Vidal}

\institute{Departament d'Astronomia i Astrof\'isica, Universitat de Val\`{e}ncia,
           c/ Dr. Moliner 50, E-46100 Burjassot, Val\`{e}ncia, Spain\\
           \email{I.Marti-Vidal@uv.es}
           \and
           Instituto de Astrof\'isica de Andaluc\'ia, CSIC,
           c/ Camino bajo de Hu\'etor 50, E-18008 Granada, Spain
           \and
           Max-Planck-Institut f\"ur Radioastronomie,
           Postfach 2024, D-53010 Bonn, Germany
           \and
           Jodrell Bank Observatory,
           Macclesfield, Cheshire SK11 9DL, United Kingdom
           \and
           Harvard-Smithsonian Center for Astrophysics,
           60 Garden St., MS 51, Cambridge, MA 02138, USA
           \and
           Department of Physics, Marquette University,
           P.O. Box 1881, Milwaukee, WI 53201-1881, USA
           \and
           National Radio Astronomy Observatory,
           P.O. Box O, Socorro, NM 87801, USA
           \and
           Naval Research Laboratory, Code 7210,
           Washington, DC 20375-5320, USA
           \and
           Department of Optics \& Quantum Electronics,
           University of Szeged, Szeged, H-6720, Hungary}

    \date{Accepted 22/05/2007}

\abstract{We report on 8.4\,GHz Very Long Baseline Interferometry
(VLBI) observations of the type II-P supernova \object{SN\,2004et} in the
spiral galaxy NGC\,6946, made on 20 February 2005 (151 days after
explosion). The Very Large Array (VLA) flux density was
1.23$\pm$0.07\,mJy, corresponding to an isotropic luminosity at
8.4\,GHz of (4.45$\pm$0.3)$\times$10$^{25}$\,erg\,s$^{-1}$\,Hz$^{-1}$ 
and a brightness temperature of (1.3$\pm$0.3)$\times$10$^{8}$\,K. 
We also provide an improved source position, accurate to about 
0.5 mas in each coordinate. The VLBI image
shows a clear asymmetry. From model fitting of the size of the
radio emission, we estimate a minimum expansion velocity of
15,700$\pm$2,000\,km\,s$^{-1}$. This velocity is more than twice
the expected mean expansion velocity estimated from a synchrotron
self-absorbed emission model, thus suggesting that synchrotron
self-absorption is not relevant for this supernova. With the
benefit of an optical spectrum obtained 12 days after explosion,
we favor an emission model which consists of two hot spots on an
underlying expanding shell of width comparable to that of
SN\,1993J.}


\maketitle

\section{Introduction}
Supernova \object{SN\,2004et} in galaxy \object{NGC\,6946} was discovered by Zwitter et al. 
(\cite{Zwitter2004}) on 27 September 2004. Yamaoka et al. (\cite{Yamaoka2004}) reported 
data from the robotic {\em T\'elescope \`{a} Action Rapide pour les Objects Transitoires} 
(TAROT) telescope from which the supernova explosion date could be well constrained. 
Nothing brighter than magnitude 19.4$\pm$1.2 was seen on September 22.017 and the supernova 
was detected at magnitude 15.17$\pm$0.16 on September 22.983. Following Yamaoka et al., 
we assume 22.0 September 2004 as the explosion date.

NGC\,6946 is a starburst spiral galaxy with a high supernova production rate. 
In fact, SN\,2004et is the eighth supernova event registered in this galaxy since 
1917. The distance estimates to NGC~6946 range
from 5.1\,Mpc (de Vaucouleurs \cite{Vaucouleurs1979}) up to 10.1\,Mpc (Sandage \& Tammann
\cite{Sandage1981}). Here we adopt 5.5$\pm$0.1\,Mpc as the distance to NGC\,6946 following
Pierce (\cite{Pierce1994}), who estimated it using the Tully-Fisher HI relation. Most of
the distance values quoted in the literature vary between 5.2 and 5.5\,Mpc. Sahu et al.
(\cite{Sahu2006}) derived a distance of 5.6\,Mpc to NGC\,6946 as an average of the 
literature values, including the result of their analysis for SN\,2004et itself, which is 
in very good agreement with the adopted 5.5\,Mpc distance.

Li et al. (\cite{Weidong2005}) found a progenitor candidate for this
supernova by analyzing {\em Canada-France-Hawaii Telescope} (CFHT) images of the field 
made before the supernova explosion.
The progenitor seems to have been a yellow supergiant that may have previously experienced a
red supergiant phase, but it could also have consisted of a binary system of red and blue
supergiants. This latter case would imply a progenitor similar to that for \object{SN\,1993J}
(Van Dyk et al. \cite{VanDyk2002}).

Spectroscopic observations of Zwitter et al. (\cite{Zwitter2004}) and Filippenko et al.
(\cite{Filippenko2004}) suggest that SN\,2004et is a Type II supernova. Hence, according
to the emission model of Chevalier (\cite{Chevalier1982b}), tested with 
other supernovae like SN\,1993J (e.~g., Marcaide et al. \cite{Marcaide1997}), we
expected that this supernova would emit detectable radio synchrotron radiation sometime after its
optical detection. Indeed, the radio emission from SN\,2004et was detected and monitored at 
5\,GHz with the {\em Multi-Element Radio Linked Interferometer Network} (MERLIN) and the VLA arrays 
from October 2004 onwards (Argo et al. \cite{Argo2005}, Kelley et al. in preparation).
Monitoring was also carried out at 6.0\,GHz and 1.6\,GHz by these authors. Unfortunately,
the monitoring observations by Argo et al. were unable to constrain the date of the 5\,GHz peak
emission, because scheduling constraints meant that no observations could be made between 3$-$17
November 2004 (see Fig. 2 of Argo et al. \cite{Argo2005}). From their observations, Argo et al.
(\cite{Argo2005}) estimated the 5\,GHz peak flux density as $\sim$2.5\,mJy in early November 2004.
Kelley et al. (in preparation), using observations of SN\,2004et at 6 different frequencies with the VLA,
fitted the flux density evolution of the supernova to the ``mini-shell'' model of Chevalier
(\cite{Chevalier1982}, \cite{Chevalier1982b}) and obtained an estimate of the SN\,2004et peak flux density 
at 5\,GHz of 2.45$\pm$0.08\,mJy on day 66$\pm$6 after explosion (assuming 22.0 September 2004 as the 
explosion date).

We arranged very-long-baseline-interferometry (VLBI) observations of
SN\,2004et in order to learn as much as possible about the radio emitting structure, the expansion
scenario, and the emission mechanisms. In Sect. 2 we describe our observations;
in Sect. 3 we present our VLBI image of SN\,2004et and the results of fits of several emission models
to the data; in Sect. 4 we discuss the magnetic field strength and the amount of energy involved
in the radio emission, together with the expansion velocity of the emitting region and the importance of
possible absorption mechanisms; in Sect. 5 we present our conclusions.

\section{Observations and data reduction}

\subsection{The VLBI data}

In our observations, made on 20 February 2005, the participating stations were the complete
Very Long Baseline Array (VLBA) (10 identical antennae of 25\,m diameter spread over the US
from the Virgin Islands to Hawaii), the phased VLA (equivalent to 130\,m diameter, New Mexico,
USA), the Green Bank Telescope (100\,m, West Virginia, USA), Goldstone (70\,m, California, USA),
Robledo (70\,m, Madrid, Spain), Effelsberg (100\,m, Bonn, Germany), and Medicina
(32\,m, Bologna, Italy).

The recording rate was 256\,Mbps, with 2-bit sampling and single polarization mode, covering
a total synthesized bandwidth of 64\,MHz (except at the phased VLA, where the
effective bandwidth was 50\,MHz). These data were correlated at the Array Operations Center 
of the National Radio Astronomy Observatory (NRAO) in Socorro (New Mexico, USA) with an averaging 
time of 2\,seconds.

The 12-hour long, 8.4\,GHz VLBI phase-reference observations were designed to maximize the
probability for the detection of supernova SN\,2004et.
We used the source \object{J2022+614} as the phase calibrator (a 3.0\,Jy quasar at an
angular distance of 2.2$^{\circ}$ from the supernova). We spent 25\% of the observing
time on the calibrator, with duty cycles of approximately 4\,minutes.
Every 22\,minutes we also observed a secondary phase calibrator,
the quasar \object{J2035+582}, with a 0.18\,Jy total flux density and an angular separation
of 1.76$^{\circ}$ from the supernova.

The correlation coefficients were then read into the NRAO Astronomical Image
Processing System ({\sc aips}) for calibration purposes. A manual phase alignment
between the 8 different IFs was obtained by fringe fitting a selected scan of the
calibrator source J2022+614.

We then performed the amplitude calibration using gain curves for each antenna
(information provided by NRAO and the European VLBI Network, EVN) and system temperatures
measured by each station during the observations.

Such a scheme is appropriate for all stations except the phased VLA. In that
case, the amplitude calibration requires previous knowledge of the flux densities
of all sources. For consistency, we first mapped the calibrators using the VLBI
data from all stations but the VLA. Then, we used the total fluxes obtained in the
mapping of J2022+614 and J2035+582 to calibrate the phased VLA.

\begin{figure}[t]
\centering
\includegraphics[width=8.25cm]{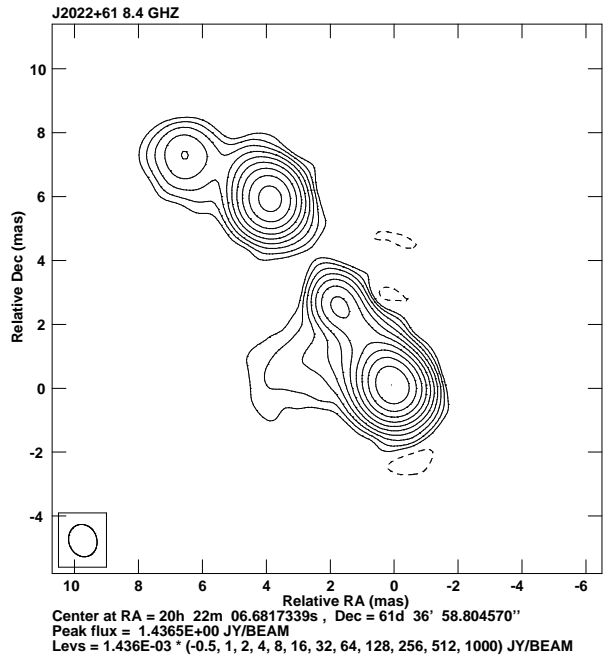}
\caption{Contour image of our phase calibrator source, J2022+614, at 8.4\,GHz. The dynamic range
(peak/rms) obtained for this image is 68,000:1. The FWHM of the CLEAN beam is shown at bottom left.}
\label{j2022}
\end{figure}

The J2022+614 data were then fringe-fitted in a standard manner and the
resulting calibrated visibilities were exported to the Caltech program {\sc difmap}
(Shepherd et al. \cite{Shepherd1995}) for imaging purposes using standard phase
(and later also amplitude) self-calibration techniques. The phase corrections derived
from self-calibration differed from the corrections derived from fringe fitting by typically
5$-$10\,degrees, and the amplitudes after self-calibration differed from the system temperature
calibrated amplitudes by typically 2$-$5\%. The resulting map of J2022+614 is shown in Fig.
\ref{j2022}. Afterwards, the J2022+614 model obtained in {\sc difmap} was read back into
{\sc aips} for a second fringe-fitting in which we obtained residual phases, delays, and
rate solutions nearly free of structure contributions.

These fringe-fitted solutions were then linearly interpolated to the times of
the SN\,2004et scans and were used for the calibration of the phases, delays and
delay rates of the VLBI SN\,2004et data.
Finally, we exported the calibrated supernova data for hybrid mapping in {\sc difmap}.

\subsection{The VLA data}

The VLA provided two kinds of data:
VLBI data from the phased array and individual antenna data. The correlation
coefficients for all the pairs of antennas could thus be obtained and
used for wide-field imaging around SN\,2004et and for obtaining an
accurate estimate of the total flux density of SN\,2004et.

\begin{figure}[t]
\centering
\includegraphics[width=8.5cm,angle=0]{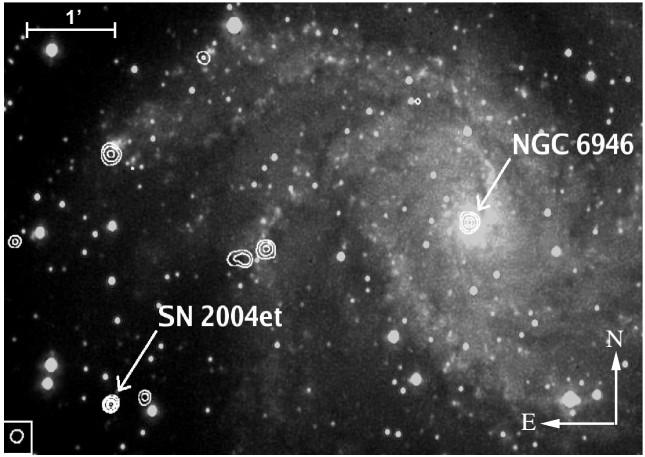}
\caption{Radio image contours from the VLA superimposed on an optical image of NGC\,6946 
(taken with the TROBAR telescope of the Observatori Astron\`omic de la Universitat de Val\`encia). The
FWHM of the CLEAN beam is shown at bottom left. Contours are at 0.02, 0.08, 0.32, 0.64, and 1 times
the peak flux density, which is 1.23 mJy/Beam.}
\label{composite3}
\end{figure}

We also used J2022+614 as the flux calibrator for the VLA data. The radio image 
contours thus obtained are shown in Fig. \ref{composite3} superimposed on an optical image.
The map is affected by bandwidth smearing effects. The radio sources identified in
Fig. \ref{composite3} had previously been listed by Hyman et al. (2000). Since the
correlation was centered on SN\,2004et, its flux density was accurately
determined as 1.23$\pm$0.07\,mJy.

\section{VLBI image of SN\,2004et}

In {\sc difmap}, we deleted the bad visibility amplitudes of SN\,2004et (i.~e., less than 0.1\%
of the data) and Fourier inverted the remaining data into the sky plane. We then applied the natural
weighting scheme to the visibilities for sensitivity optimization and detected a nearly point
emitter (showing a small elongation in the South-East direction) whose peak was displaced 5.1\,mas
West and 14.7\,mas South from the coordinates at which we had centered the correlation (we used the
coordinates given by Beswick et al. \cite{Beswick2004}). The emitter is undoubtedly SN\,2004et. 
Thus, a better position of the supernova, based on a phase-reference to J2022+614, a source of the
{\em International Celestial Reference Frame} (ICRF), is:
$\alpha$ = 20$^{\rm h}$\,35$^{\rm m}$\,25.3596$^{\rm s}$ and $\delta$ = +60$^{\circ}$ 07' 17.7203"
(equinox J2000.0, $\pm$0.5\,mas\footnote{This error corresponds to the beamwidth of the VLBI interferometer
applying the natural weighting scheme to the SN\,2004et visibilities.} in each coordinate).

\begin{figure}[t]
\centering
\includegraphics[width=8.25cm,angle=0]{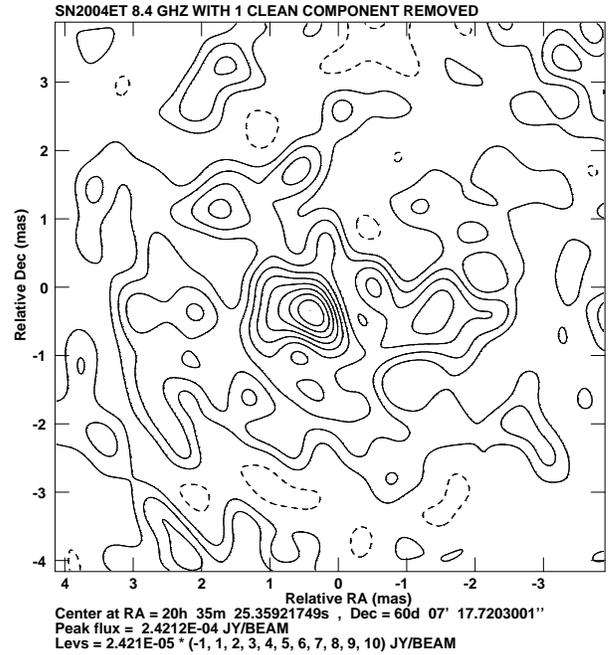}
\caption{Residuals after the subtraction of a single point source to SN\,2004et (see text in Sect. 3).}
\label{sn2004et-1clean}
\end{figure}

Given the very low flux-density of SN\,2004et, we did not apply any
phase self-calibration to the data.
We fitted several emission models to the real and imaginary parts
of the visibilities using the {\sc difmap} task {\em modelfit}.
The errors of the fitted parameters of the models correspond to the
statistical uncertainties taken from the covariance matrices of
the fits.
A point source model fitted to the data has a total flux density
of 0.63$\pm$0.03\,mJy. After subtraction of this point source model
from the data, there is still clear extra emission
at the level of 0.24\,mJy\,beam$^{-1}$, indicating structure in the
supernova emission. A map of this extra emission is shown in Fig.
\ref{sn2004et-1clean}.

\begin{figure}[t]
\centering
\includegraphics[width=8.25cm,angle=0]{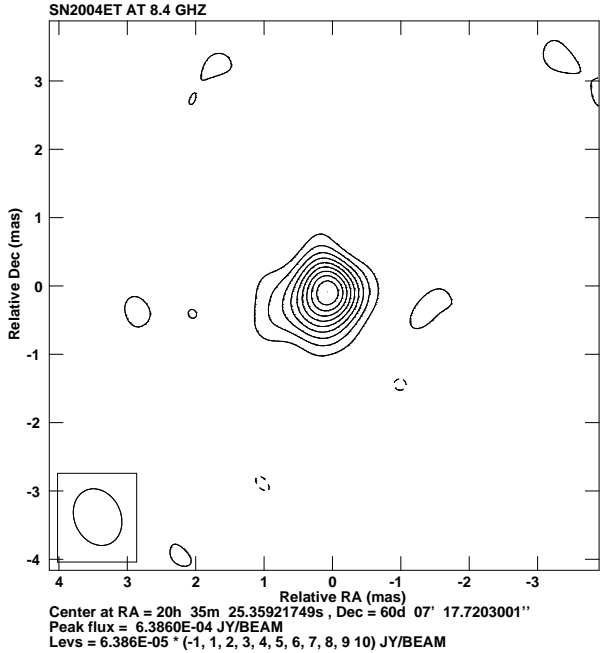}
\caption{Contour map of SN\,2004et obtained from our VLBI data by model fitting two
point sources (see text in Sect. 3). The lowest positive level shown is 10\% of
the peak, which is 0.64\,mJy/Beam. The FWHM of the CLEAN beam is shown at bottom left.}
\label{sn2004et-vlbi}
\end{figure}

This extra emission could be due to a bad antenna calibration, so we checked on
this possibility by making a map of the source leaving out one antenna of the array at a time.
We also imaged the first and second halves of these data separately to check the structure
repeatability.
All the resulting maps showed the extra emission when we subtracted a point source from
the uv-data. Thus, we conclude that this emission belongs to SN\,2004et and is
not produced by imperfect calibration.

Using two point emitters to fit the data, we recovered
a flux density of (0.57$\pm$0.05)\,mJy for the first component and (0.26$\pm$0.05)\,mJy for
the second, located at an angular distance of 0.5$\pm$0.06\,mas and at a position angle of
(136$\pm$4)$^{\circ}$ with respect to the strongest component.

Given that the total recovered flux density in the uv data using
{\em CLEAN} adds up to 0.87\,mJy (5\% higher than the flux density
fitted), and the VLA flux density estimate is 1.23\,mJy, there
must be some missing flux density in our VLBI data. Most of
this missing flux density could be due to a loss of coherence in the
phase-referenced VLBI data compared to the VLA data. This part of
the missing flux could be recovered using phase self-calibration,
were the flux density of the source high enough. However, given
that this is not the case, we decided to leave the phases
unchanged and consider this lack of flux density as an
instrumental limitation.

Disks or shell-like sources could also be used to fit the data. Indeed, we show the results
of some of these fits in Table \ref{TablePaper}. We find that for the fit to two point emitters the
resulting rms of the residuals is slightly (but not significantly) smaller than for any of the
other fits, which are very similar. The similarity between the rms of the fits of all the different
emission models indicates that the source is only barely resolved by the interferometer and it is not
possible to distinguish between more detailed models.

We show the VLBI image of SN\,2004et in Fig. \ref{sn2004et-vlbi}. The peak brightness is
0.64\,mJy beam$^{-1}$ and the rms of the residuals 0.03\,mJy\,beam$^{-1}$.

\begin{table*}
\caption{Results of different models fit to the SN\,2004et data.\newline
$^{1}$ The assumed shell width is 30\% of the outer radius.\newline
$^{2}$ The diameter refers to the separation of the point emitters.}
\label{TablePaper}
\centering
\begin{tabular}{c c c c}
\hline\hline
\textbf{Model} &\multicolumn{3}{c}{\textbf{Results}}\\
 & Diameter & Total flux density & rms of residuals\\
 & (mas) & (mJy) &  ($\mu$Jy/beam)\\
\hline
Uniform disk & 0.62 $\pm$ 0.08 & 0.81 $\pm$ 0.05 &  27 \\
Optically thin sphere & 0.8 $\pm$ 0.1 & 0.83 $\pm$ 0.05 & 27 \\
Optically thin shell$^{1}$ & 0.70 $\pm$ 0.08 & 0.82 $\pm$ 0.05 & 27 \\
Two point emitters$^{2}$ & 0.50 $\pm$ 0.06 & 0.83 $\pm$ 0.07 & 26 \\
\hline
\end{tabular}
\end{table*}

\section{Discussion}

\subsection{Supernova energy and magnetic field}

The SN\,2004et flux density measured by the VLA on February 2005 using J2022+614 as the flux
density calibrator is 1.23$\pm$0.07\,mJy. For a distance to NGC\,6946 of
5.5$\pm$0.1\,Mpc, this flux density corresponds to an isotropic luminosity at
8.4\,GHz of (4.5$\pm$0.3)$\times$10$^{25}$\,erg\,s$^{-1}$\,Hz$^{-1}$. The corresponding
brightness temperature, if we use the size determined from the fit to two point emitters, is
(1.3$\pm$0.3)$\times$10$^{8}$\,K, indicating a non-thermal synchrotron origin of the
radio emission.

We did not obtain spectral information of the radio emission coming
from SN\,2004et in our VLBI observations. However, Kelley et al. (in preparation) observed
SN\,2004et with the VLA at several frequencies (ranging from 1.4\,GHz to 43.3\,GHz) during several
hundred of days following the supernova explosion. These authors fit the evolution of the resulting
radio light curves to a given parametric model, from which they estimate a spectral index,
$\alpha = -0.87 \pm 0.04 $, for the radio emission in the optically thin phase of the supernova
evolution. Given that the supernova was in its optically thin phase during our VLBI observations, we
used the spectral index estimated by Kelley et al. From the previously derived
luminosity at 8.4\,GHz, we estimate a total integrated radio luminosity of
(1.3$\pm$0.1)$\times10^{36}$\,erg\,s$^{-1}$ between 1.4\,GHz and 43.3\,GHz.

Using this result together with the SN\,2004et size estimate from our VLBI
observations, we can derive the minimum total energy in the radiating region of the
supernova and the corresponding equipartition magnetic field.
From Pacholczyk (\cite{Pacholczyk1970}) (see chapter 7, formulae 14 and 15) we have:

\begin{equation}
E_\mathrm{min} = c_{13}(1 + k)^{4/7}\phi^{3/7}R^{9/7}L_{R}^{4/7}
\label{EquipE}
\end{equation}

\begin{equation}
B_\mathrm{eq} = \left (4.5 c_{12} (1 + k)/{\phi} \right )^{2/7}R^{-6/7}L_{R}^{2/7}
\label{EquipB}
\end{equation}

\noindent where $c_{12}$ and $c_{13}$ are functions of the spectral index, $\alpha$, and the
minimum and maximum frequencies considered in the spectrum integration, $\phi$ is the filling
factor of the emitting region, $R$ is the source radius, $L_{R}$ is the integrated radio luminosity,
and $k$ is the ratio between the heavy particle energy and the electron energy.
The mechanism of generation is not well known, hence $k$ can vary from 1 to
$m_{p}/m_{e} \approx 2\times10^{3}$. 
Thus, these formulae give us a range of possible values of the equipartition energy and
magnetic field instead of an estimate. Also we do not know the
value of the filling factor, $\phi$ (the fact that the emission may not be completely
isotropic will also affect the estimate), but the effect of $\phi$ on the value of
$E_\mathrm{min}$ and $B_\mathrm{eq}$ goes aproximately as the square root and the cube root,
respectively, as can be seen in equations \ref{EquipE} and \ref{EquipB}.

Assuming that the radio emitting region corresponds to a shell of
width equal to 30\% of the outer radius (e.~g., Marcaide et al.
\cite{Marcaide1995}, \cite{Marcaide2007}), we find that the filling
factor, $\phi$, is 0.66, and that the minimum total energy (particles
plus fields) of the emitting region (with a radius of
0.70$\pm$0.08\,mas) ranges between 8.2$\times$10$^{45}$\,erg and
4.22$\times$10$^{47}$\,erg. The upper part of the range still
represents less than one part in a thousand of the supernova
emission energy, $\sim$10$^{51}$\,erg.

The corresponding equipartition magnetic field ranges between
37\,mG and 260\,mG. This range is similar to those obtained for
other type II supernovae, like \object{SN\,1986J} (P\'erez-Torres et al.
\cite{PerezTorres2002}), SN\,1993J (e.~g., P\'erez-Torres et al.
\cite{PerezTorres2001}) or, more recently, \object{SN\,2001gd}
(P\'erez-Torres et al. \cite{PerezTorres2005}). Values in this
range are very high compared to the typical (about 1\,mG) magnetic
fields in the circumstellar medium of a massive star. Given that
the hydrodynamic compression from the supernova shock only
increases the circumstellar magnetic field by a factor of about 4
(e.\,g., Dyson \& Williams \cite{Dyson1980}), other field
amplification mechanisms, i.~e. turbulent amplification possibly
due to Rayleigh-Taylor instabilities (Chevalier
\cite{Chevalier1982b}), must be taking place in the emitting region
of SN\,2004et.

\subsection{Supernova structure}

Our results show that the emission structure of supernova SN\,2004et is
not spherically symmetric. The residuals of the fit of a point emitter to the VLBI data,
shown in Fig. \ref{sn2004et-1clean}, indicate that the emitting structure
has an angular asymmetry.

The structure of the ejecta of type II-P supernovae are supposed to be very
spherical, given that their extended hydrogen envelope should ``spherize'' the possible
asymmetries in the shock (e.~g., Khokhlov, H\"oflich \& Wang \cite{Khokhlov2001}).
Thus, the asymmetry found in SN\,2004et is likely to come from the structure of
the circumstellar material.

Assuming a circumstellar origin for the asymmetry of the SN\,2004et emission, we may
interpret such asymmetry in two ways:

{\em Model 1.} A shell with two hot spots, one on each ``side'' of the emission of the shell.
These hot spots could be due to anisotropies
of the circumstellar density distribution (probably due to asymmetries in the progenitor
pre-supernova wind) and/or anisotropies of the magnetic field. This
kind of anisotropic shell has been found in the case of SN\,1986J
(Bartel et al. \cite{Bartel1991}, P\'erez-Torres et al. \cite{PerezTorres2002},
Bietenholz et al. \cite{Bietenholz2004}). The fact that we recovered a much lower flux density
with VLBI than with the VLA is consistent with this model. The two hot spots
of the radio structure would emit the major part of the total flux density of the source,
and the emission from the shell, being faint, would be buried in
the thermal noise of the map. A flux-density distribution with a peak surface brightness just
under 0.1\,mJy\,beam$^{-1}$ (peak flux density of the noise; see the spread contours
of Fig. \ref{sn2004et-1clean}) distributed over an area of 2 VLBI beams (roughly, the area of the
radio structure), would account for a flux density of $\sim$0.2\,mJy, that is, half of the
difference between the flux density recovered by the VLA and VLBI.

{\em Model 2.} An expanding shell plus a protrusion. This protrusion could be produced by
an asymmetry in the circumstellar medium distribution, like an axi-symmetric region with a
lower circumstellar density. Such asymmetry could result in a faster expansion of the portion
of the forward shock located in that region.
According to hydrodynamical simulations (Blondin, Lundqvist, \& Chevalier \cite{Blondin1996}),
the length of the protrusion could be more than twice the radius of the main shell, but the
time required for the formation of such a structure in the forward shock would be of the order
of several years, much larger than the age of SN\,2004et at the time of our observations.

Given the youth of the supernova (151\,days at the epoch of our observations), the size of the
protrusion should be much smaller than the mean size of the
shell. This argument does not favour {\em Model 2} as the origin of the
observed structure.

\subsection{Expansion velocity}

The different models shown in Table \ref{TablePaper} yield different estimates
of the size of SN\,2004et. We do not know which one of these models (or another) best
represents the true emitting structure of SN\,2004et. Thus, we will estimate the minimum
expansion velocity compatible with our data using the smallest of our size estimates. This
separation (see Table \ref{TablePaper}) is 0.50$\pm$0.06\,mas and corresponds to the model
of two compact emitters.

Assuming a distance of 5.5$\pm$0.1\,Mpc for NGC\,6946 and 22.0 September 2004 as the
explosion date of the supernova, we find that the radial distance between the point
components translates into a mean separation speed of
31,400$\pm$4,000\,km\,s$^{-1}$ between them. Assuming also that the two compact
emitters come from two diametrically opposed hot spots on an expanding shell, the center
of the explosion would fall half way between them and the expansion velocity of the shell would be
15,700$\pm$2,000\,km\,s$^{-1}$. This velocity is high compared with the theoretical predictions
of a synchrotron self-absorbed model, as will be shown below.

If we use the size estimates of the other models in Table \ref{TablePaper}, the derived
expansion velocities are higher than 15,700\,km\,s$^{-1}$. Fitting the data to a disk with uniform
emission (unrealistic, because the circumstellar region is not optically thick at this
epoch), yields an expansion speed of the ejecta of 19,500$\pm$2,500\,km\,s$^{-1}$;
fitting to an optically thin shell-like model yields a speed 22,000$\pm$2,500\,km\,s$^{-1}$, and
fitting to an optically thin filled sphere results in a speed of 25,200$\pm$3,000\,km\,s$^{-1}$.

\begin{figure}[t]
\centering
\includegraphics[width=8.25cm]{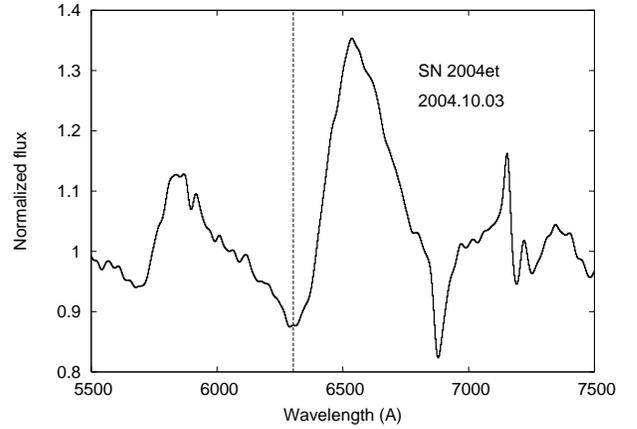}
\caption{Red part of an optical spectrum of SN\,2004et obtained at
the David Dunlap Observatory (see Vinko et al., in preparation). The
spectrum is normalized to the continuum. The vertical dashed line marks
the position of the blue absorption minimum of H$\alpha$, corresponding
to the velocity of the H$\alpha$ ``photosphere'' (see text).} \label{vinko}
\end{figure}

An early optical spectrum of SN\,2004et is shown in Fig.
\ref{vinko} and shows characteristic P Cygni profiles of an
expanding envelope. This spectrum was taken on 3 October 2004 (7
days after discovery and 12 days after explosion) at the David
Dunlap Observatory in Canada with the Cassegrain spectrograph on
the 74\,inch (188\,cm) telescope as part of observations that will
be fully described by Vinko et al. (in preparation). From the
position of the blue absorption minimum of the H$\alpha$ line, an
expansion velocity of the line of $11,900\pm600$\,km\,s$^{-1}$ is
estimated, once the redshift of the host galaxy is corrected. This
velocity represents the velocity of the H$\alpha$
``photosphere'', i.~e. where the optical depth becomes $\tau \sim
1$. Because at early phases the ejecta are almost fully ionized,
this layer should be close to the top of the ejecta. Thus, the
velocity from the H$\alpha$ line should also be close to
(although somewhat lower than) the expansion velocity of the
ejecta.

Sahu et al. (\cite{Sahu2006}) have reported expansion speeds from
optical observations starting on day 30 after explosion (25 days
after discovery). They identify a low velocity component in
H$\alpha$, which is the same as that shown in Fig. \ref{vinko}
but for later days and hence at lower velocities. They also
identify a high velocity component for days 30 through 90. Such
high velocity feature is expected to arise from the interaction
between the CSM and the SN shock wave (Chugai et al.
\cite{Chugai2007}). It corresponds to a feature placed on the
absorption side of the P Cygni profile of the lower velocity
H$\alpha$, both in their spectra and in the spectrum shown in
Fig. \ref{vinko}. However, this component appears very weak and
almost imperceptible in Fig. \ref{vinko}. We are not convinced
that it provides a good estimate of the early expansion speed and
we will not use hereafter. Also, Chugai et al. (\cite{Chugai2007})
did not find convincing evidence for the high-velocity
H$\alpha$ feature in the middle and late plateau phases, when
such features should be more pronounced.

According to the model of Chevalier (\cite{Chevalier1982b}), the
optical emission from young supernovae originates in the shocked
ejecta region, located behind the shocked circumstellar region.
This implies that the maximum velocities inferred from the optical
lines should always be smaller than the velocities inferred from
the VLBI observations of the radio emitting structure. The latter
radio emission would originate in the shocked circumstellar
region. In fact, according to the self-similar expansion model of
Chevalier (\cite{Chevalier1982}, \cite{Chevalier1982b}), the ratio 
between the maximum
velocity inferred from the optical spectral lines in the very
early days and the expansion velocity of the radio shell should be
similar to the ratio between the inner and the outer radii of the
shell-like radio emitting structure. Marcaide et al.
(\cite{Marcaide2007}) find a ratio of 0.69$\pm$0.02 between the
inner and the outer radii of the shell-like structure of supernova
SN\,1993J.

For SN\,2004et we find that the ratio between the maximum 
velocity inferred from the optical spectral lines in the very early 
days and the expansion velocity of the radio shell is about 0.76 for
the model yielding the smallest expansion speeds from the radio
data. Thus, if the components of the two-point model are hot spots
on the outer part of an underlying shell (which we do not detect,
as explained above, due to its faintness) the width of such a shell
would be comparable to that of SN\,1993J. In arriving at this 
conclusion, we have not considered the deceleration of the supernova 
ejecta between the epoch of the optical observations and the epoch of 
the VLBI observations. This neglect implies that the optical line 
velocity at our VLBI epoch could be lower than 11,900\,km\,s$^{-1}$. 
Thus, also taking into account that 15,700\,km\,s$^{-1}$ 
is the smallest expansion speed estimated from these radio data, we
conclude that the real ratio between the inner and the outer radii 
of the supernova radio shell could be somewhat smaller than 0.76.

For cases of supernovae where synchrotron self-absorption (SSA) dominates over the circumstellar
free-free absorption, and where there is equipartition of the energy between relativistic
particles and magnetic fields, Chevalier (\cite{Chevalier1998}) found a relationship between the
observing frequency, the size of the emitting region, the peak intensity of the supernova light curve,
the supernova age at the peak, and the spectral index of the energy distribution of the relativistic
electrons. For such a case, one can estimate the expansion velocity of a supernova based on
measurements of the peak luminosity at a given frequency, the supernova age at the peak, and the
synchrotron spectral index, as follows:

\begin{eqnarray}
V_\mathrm{SSA}(\mathrm{km}\,\mathrm{s}^{-1}) = 5.3786\times10^{6} ({\beta\phi})^{-0.0541}
\left (\frac{F_{\mathrm{p}}}{1 \mathrm{Jy}}\right )^{0.4729}\times \nonumber \\
\left (\frac{D}{1 \mathrm{ Mpc}}\right )^{0.9458} \left (\frac{\nu}{1.0 \mathrm{ GHz}}\right )^{-1}
\left (\frac{t_{p}}{1 \mathrm{ day}}\right)^{-1}
\label{VelSSA}
\end{eqnarray}

\noindent where $\beta$ is the ratio of the relativistic particle
energy density to the magnetic field energy density (since we
assume energy equipartition, $\beta = \frac{4}{3(1+k)}$,
Pacholczyk \cite{Pacholczyk1970}, chapter 7), $\phi$ is the
filling factor of the emitting region for a sphere (we assume a
shell of width equal to 30\% of the outer radius, which yields
$\phi = 0.66$), $F_{\mathrm{p}}$ is the flux density at the peak, $D$ is
the distance, $\nu$ is the observing frequency, and $t_{p}$ is the
age at the peak. We use the synchrotron spectral index found
by Kelley et al. (in preparation) to set the index of the electron
energy distribution to $\gamma = 2.74$ (the energy distribution of
the relativistic electrons is assumed to be of the form $N \propto
E^{-\gamma}$ and $\gamma = 1 - 2\alpha$).

When we use equation \ref{VelSSA} and assume for $F_{p}$ the 5\,GHz peak flux density published by Argo
et al. (\cite{Argo2005}) (i.\,e., $F_{p} =$ 2.5\,mJy at $t_{p} =$ 42\,days), we estimate a SSA expansion
velocity of the radio shell of SN\,2004et of 9,700$\pm$1,800\,km\,s$^{-1}$. This is much lower 
than the estimate for each of the models in Table \ref{TablePaper}. If instead we use the 5\,GHz peak
flux density fitted by Kelley et al. (in preparation) at day 66$\pm$6 after explosion, which appears more
accurate given the extended time coverage of their observations near the emission peak at 5\,GHz, the
SSA expansion velocity of the radio shell is only 6,100$\pm$1,100\,km\,s$^{-1}$. This result is less 
than half of any of the estimates based on our VLBI data.

As noted by Chevalier (\cite{Chevalier1998}), if the free-free
absorption from the circumstellar medium is comparable to, or even
higher than, the synchrotron self-absorption, the peak of emission
will be defined by the free-free absorption evolution. This will
produce a negative bias in the estimate of the ejecta expansion
velocity, as we seem to find for SN\,2004et. Comparison of VLBI
results and radio light curves for other supernovae also gives
important negative biases (see Chevalier \cite{Chevalier1998}),
but at present there is an insufficient number of cases (in fact,
only SN\,1979C, SN\,1986J, and SN\,1993J) to reach firm
conclusions about the typical values of these biases. For
supernovae SN\,1986J and SN\,1979C, the bias ratios $V_{{\rm
VLBI}}/V_{{\rm SSA}}$ ($\sim$2.56 and $\sim$2.60, respectively)
are similar to that for SN\,2004et (if we use the minimum size
compatible with our data), but for SN\,1993J this bias ratio is
smaller ($\sim$1.5).

\subsection{Comparison with other supernovae in NGC\,6946}

From those eight supernovae known to have exploded in NGC\,6946, only 
four have been detected as radio supernovae: \object{SN\,1968D}, \object{SN\,1980K},
\object{SN\,2002hh}, and SN2004\,et. For two (SN\,1980K and 
SN\,20004et), it has been possible to fit the light curves (Weiler et al. 
\cite{Weiler1992}, Kelley et al. in preparation) in terms of the ``mini-shell''
model (Chevalier \cite{Chevalier1982}, \cite{Chevalier1982b}).

There are a number of similarities between SN\,1980K and SN\,2004et. 
The flux-density power-law time decay parameter takes the values $-$0.62 and $-$0.74, 
respectively, and the power-law parameter of the evolution of the external
opacity with time takes the values $-$2.46 and $-$3.07,
respectively. But there are also significant differences between
these supernovae: the spectral index is $-$0.52 for SN\,1980K,
while for SN\,2004et it is $-$0.87; the time after explosion
needed to reach the peak spectral luminosity was 134$\pm$2 days
for SN\,1980K and is only 66$\pm$6 days for SN\,2004et. For the
case of SN\,1980K (Weiler et al. \cite{Weiler1992}), internal and
clumpy opacity contributions are not required to obtain a good
fit, and the light curves are well characterized by an optically
thin synchrotron radio emission absorbed by an uniform external
medium\footnote{For SN\,1968D, it has not been possible to fit the
light curves due to the late detection (at an age of $\sim$26
years), but the measured flux densities are consistent with the
extrapolated radio behaviour of SN\,1980K.}. However, for
SN\,2004et (see Kelley et al. in preparation), the absorption
produced by a uniform external medium cannot fit the data, but a
good fit is obtained for absorption by a clumpy or filamentary
external medium. The presence of this clumpy external medium could
explain the asymmetry in the radio structure that we have found in
our VLBI image. On the one hand, an expansion into a clumpy medium would yield an irregular
interaction between the ejecta and the circumstellar medium and, thus, the
evolution of the expanding material would be asymmetric. On this account, {\em Model 2}
would appear supported. On the other hand, the synchrotron emission, being proportional to the
relativistic electron density in the circumstellar region, would be larger in the clumps. Thus,
these clumps in the CSM would give rise to several hot spots in the expanding radio shell and
would appear to give support to {\em Model 1}. The asymmetry found in our VLBI image is, then,
consistent with the expansion of the supernova ejecta into a clumpy medium.

\section{Conclusions}

We report on 8.4\,GHz VLBI observations of supernova SN\,2004et in NGC\,6946.
The observations, made 151 days after explosion, are the first VLBI observations of a type II-P
supernova, and successfully detect the radio emission and yield an image of the source. This image 
shows evidence of structure.

This structure has a geometry that could be interpreted either as
emission coming from a shell with two hot spots ({\em Model 1}) or
emission coming from a region expanding anisotropically and
developing a protrusion ({\em Model 2}). However, such a protrusion
would be too large for so young a supernova, to be compatible with
the hydrodynamic simulations  of Blondin, Lundqvist, \& Chevalier
(\cite{Blondin1996}).

The asymmetry found in our VLBI image can be explained if the supernova expands in the clumpy
circumstellar medium suggested by the fits to the radio light curves of this supernova (Kelley
et al. in preparation).

The minimum expansion speed that we obtain is
15,700$\pm$2,000\,km\,s$^{-1}$ for a two point emitter model.
Using other models yields larger expansion speeds. Maximum
velocity estimates obtained from optical spectral line
observations give lower values than the velocities inferred from
our VLBI data for all cases. Such differences between maximum
optical and radio expansion speeds can be understood in the
framework of the self-similar expansion model of Chevalier
(\cite{Chevalier1982}, \cite{Chevalier1982b}). In particular, the ratio of the maximum
early $H\alpha$ line velocity (corresponding in the Chevalier
model to the maximum velocity of the ejecta) to the expansion
velocity estimated from VLBI for a two point model is $\sim$0.76,
similar to that for SN\,1993J. This ratio suggests that {\em Model
1} is a plausible model.

In any case, the expansion speed derived from our VLBI data is
more than twice larger than the estimate obtained using the
synchrotron self-absorbed model proposed by Chevalier
(\cite{Chevalier1998}). This discrepancy indicates that the
synchrotron self-absorption must be negligible (i.~e., small or
nonexistent) compared to the circumstellar free-free absorption in
SN\,2004et at our observing epoch.

The flux density obtained from our VLA data (1.23\,mJy) and our 
minimum VLBI size estimate (diameter of 0.5\,mas) translate into 
a brightness temperature of (1.3$\pm$0.3)$\times$10$^{8}$K, indicating 
a synchrotron origin for the radiation. Using the spectral index
$\alpha = -0.87 \pm 0.04$, determined by Kelley et al. (in preparation) 
with data between 1.4 and 43.3\,GHz, we find that the amount of energy 
injected into the emitting region is in the range 
(8$-$420)$\times$10$^{45}$\,erg, a small fraction (lower than 0.001) of 
the typical emission energies in supernova explosions. The 
equipartition magnetic field falls within (37$-$260)\,mG, indicating 
the existence of highly efficient field amplification due perhaps to 
turbulent instabilities in the emitting region.

Further VLBI observations of this supernova are not possible given 
its low flux density, now well below the VLBI sensitivity limit at the
possible data rates at present. VLBI 
observations of similar, Type II-P, supernovae will be essential for the 
understanding of the details of the radio emission of this kind of 
explosion.

\begin{acknowledgements}
This work has been partially funded by Grants AYA2004-22045-E,
AYA2005-08561-C03, and AYA2006-14986-C02 of the Spanish DGICYT.
The National Radio Astronomy Observatory is a facility of the
National Science Foundation operated under cooperative agreement
by Associated Universities, Inc. The European VLBI Network is a
joint facility of European, Chinese, South African and other radio
astronomy institutes funded by their national research councils.
K.~W.~W. thanks the Office of Naval Research for the 6.1 funding
supporting this research. J.~V. received support from Hungarian
OTKA Grants No. TS 049872 and T042509. We thank the
Observatori Astron\`{o}mic of the UV for obtaining the optical
image of NGC\,6946. We thank the anonymous referee for his/her
useful comments.
\end{acknowledgements}

\end{document}